\documentclass{aa}  
\usepackage{graphicx,natbib}
\usepackage{txfonts}

\graphicspath{{figs/},{./}}

\newcommand{\changes}{}

\newcommand{\citen}[1]{\citeauthor{#1} \citeyear{#1}}

\newcommand{\scri}{\scriptsize}

\def\pa{\partial}
\def\Bb{{\bf B}}
\def\Bp{B_{\phi}}
\def\rin{\frac{1}{r}}
\def\ald{\dot{F}}
\def\Rs{R_{\odot}}
\def\Beq{B_{\mbox{\scri eq}}}

\begin{document}
   \title{Development of twist in an emerging magnetic flux tube by poloidal
   field accretion}


   \author{P. Chatterjee\inst{1}
          \and
	  A. R. Choudhuri\inst{1,2}
          \and
          K. Petrovay\inst{1,2,3}
          }


   \institute{\inst{1}Department of Physics, Indian Institute of Science, 
	      Bangalore 560012, India\\
              \inst{2}E\"otv\"os University, Department~of Astronomy, 
              Budapest, Pf.~32, H-1518 Hungary\\
	      \inst{3}
              Department of Applied Mathematics, University of Sheffield, 
	      Hicks Building, Hounsfield Road, Sheffield S3 7RH, U.K. \\
              \email{K.Petrovay@astro.elte.hu}
             }

   \date{Received ; accepted}

\authorrunning{Chatterjee et al.}
\titlerunning{Development of twist in a flux tube by poloidal
   field accretion}


\abstract{}{
Following an earlier proposal by Choudhuri (2003) for
the origin of twist in the magnetic fields of solar active regions, we model
the penetration of a wrapped up background poloidal field into a  toroidal
magnetic flux tube rising through the solar convective zone. 
}{
The rise of the straight, cylindrical flux tube is followed by numerically
solving the induction equation in a comoving Lagrangian frame, while an
external poloidal magnetic field is assumed to be radially advected onto the
tube with a speed corresponding to the rise velocity.
}{
One prediction of our model is the existence of a ring of reverse current
helicity on the periphery of active regions. On the other hand, the amplitude
of the resulting twist depends sensitively on the assumed structure (diffuse
vs. concentrated/intermittent) of the active region magnetic field right before
its emergence, and on the assumed vertical profile of the poloidal field.
Nevertheless,  in the model with the most plausible choice of assumptions a
mean twist comparable to the observations results.
}{
Our results indicate that the contribution  of this mechanism to the twist can
be quite significant, and under favourable circumstances it can potentially
account for most of the current helicity observed in active regions. 
}

   \maketitle
%

\section{Introduction}

The magnetic field of a typical sunspot is approximately vertical below the
photosphere and then spreads radially at the photospheric level.  However, even
visual observations show that the penumbral structures of sunspots are often
twisted, as first pointed out by \citet{Hale:vortices} and
\citet{Richardson:vortices}.  Vector magnetogram measurements in the past
decade  have established that sunspot magnetic fields have helical structures,
with a higher occurrence of negative helicity in the northern hemisphere
(\citen{Seehafer:helicity}; \citen{Pevtsov1995}, \citeyear{Pevtsov2001};
\citen{Bao+Zhang:helicity}). One possible theoretical explanation of the
observed helicity was proposed by \citet{Choudhuri:tube.dynamo}, who suggested
that the poloidal flux in the solar convection zone (SCZ) gets wrapped around a
rising flux tube.   \citet{Choudhuri+:helicity} carried out a numerical
simulation demonstrating that the observed hemispheric handedness rule can
indeed be reproduced by incorporating this mechanism in their dynamo model 
(\citen{Nandy+Choudhuri2002}; \citen{Chatterjee2004};
\citen{Choudhuri+:reply}).

If the magnetic flux in the rising flux tube is nearly frozen, then we expect
that the poloidal flux collected by it during its rise through the SCZ would be
confined in a narrow sheath at its outer periphery. In order to produce a twist
in the flux tube, the poloidal field needs to diffuse from the sheath into the
tube by turbulent diffusion. However, turbulent diffusion is strongly
suppressed by the magnetic field in the tube. This nonlinear diffusion process
was studied in an untwisted flux tube by  \citet{Petrovay+FMI:erosion} who
concluded that a substantial amount of flux may be eroded away from a rising
flux tube during the process of its rise through the SCZ.  The model was
subsequently successfully applied for sunspot decay
(\citen{Petrovay+vDG:decay1}). In the present paper we extend this model by
including the poloidal component of the magnetic field (i.e.\ the field which
gets wrapped around the flux tube) and study the evolution of the magnetic
field in the rising flux tube, as it keeps collecting more poloidal flux during
its rise and as turbulent diffusion keeps acting on it.  

The conclusions drawn from our model hinge on some assumptions, especially
concerning the subsurface magnetic field structure in the last phases of the
rise of the tube.  Allowing the various parameters of the model to vary over
reasonable ranges, we find that the poloidal field wrapped around the flux tube
should be able to penetrate inside the flux tube if the magnetic field falls
below the equipartition value in the top layers of the convection zone. On the
other hand, if some physical effects keep the magnetic field well above the
equipartition value, then the poloidal field may remain confined in a sheath
instead of penetrating to the core of the flux tube.  We expect that much more
high-quality magnetogram data will be available in the future and more will be
known as to how the current helicity in a sunspot and in an active region
varies with radial distance from the centre.  This may enable detailed
comparisons between theory and observations in future, with the possibility of
constraining various parameters in the theory.  Our aim at present is only to
set up the basic theoretical framework and study some exploratory solutions. 

Section~2 presents the mathematical formulation of our problem, while the 
numerical solutions are presented and discussed in Section~3. Section~4
concludes the paper.

\section{Mathematical formulation}

We consider a straight, cylindrical, horizontal magnetic flux tube rising
through the solar convective zone. As all variables in this model depend only
on the radial distance from the tube axis and on time, we will study the
wrapping of the large-scale poloidal field around the flux tube by considering
a radially symmetric accretion of azimuthal field by the flux tube. A further
complication is the expansion of the flux tube during its rise, due to the
decrease of the external pressure. This expansion is assumed to be
self-similar. Certainly self-similarity is not a bad assumption. We expect the
tube to expand in such a way that the density inside the tube remains
homogeneous. A self-similar expansion ensures that.  

\subsection{Equations in the comoving Lagrangian frame}

Suppose we formulate our problem in a frame of reference fixed with the centre
of the rising flux tube.  
Substituting a radial expansion velocity
$${\bf v} = v {\bf e}_r\eqno(1)$$
and an axisymmetric twisted magnetic field
$${\bf B} = B_z {\bf e}_z + B_{\phi} {\bf e}_{\phi}\eqno(2)$$
in the induction equation
$$\frac{\pa \Bb}{\pa t} = \nabla \times ({\bf v} \times \Bb)
- \nabla \times (\eta \nabla \times \Bb), \eqno(3)$$
we get the following equations for the axial and poloidal fields in the flux
tube:
$$\frac{\pa B_z}{\pa t} + \rin \frac{\pa}{\pa r}(rvB_z)
= \rin \frac{\pa}{\pa r} \left( \eta r \frac{\pa B_z}{\pa r} \right), 
\eqno(4)$$
$$\frac{\pa \Bp}{\pa t} + \frac{\pa}{\pa r}(v \Bp)
= \frac{\pa}{\pa r} \left[ \eta \rin \frac{\pa}{\pa r}(r \Bp) \right].
\eqno(5)$$

We first consider what happens in the interior of the flux tube as it expands
and is subject to turbulent diffusion.  We shall discuss later how accumulation
of additional poloidal flux during its rise can be incorporated. The
independent variables in the two equations (4) and (5) are $r$ and $t$. Let us
assume that the material inside the flux tube expands in a self-similar fashion
and  use the Lagrangian position coordinate $\xi$ of a fluid element instead of
$r$.  The initial value of $r$ at time $t=0$ can be taken as the value of
$\xi$.  Then we can write
$$\xi = F (t) r, \eqno(6)$$
where $F (t)$ will have to be a monotonically decreasing function of $t$ for an
expanding flux tube.  The assumption of self-similarity implies that we have to
use the same factor  $F (t)$ for all fluid elements inside the flux tube to go
from the radial coordinate $r$ to the Lagrangian coordinate $\xi$. In equations
(4) and (5) we now want to transform from variables $(r,t)$ to $(\xi, t)$.  By
the chain rule of partial differentiation, it can easily be shown that
$$\left( \frac{\pa}{\pa r} \right)_t = F 
\left( \frac{\pa}{\pa \xi} \right)_t, \eqno(7)$$
$$\left( \frac{\pa}{\pa t} \right)_r = 
\left( \frac{\pa}{\pa t} \right)_{\xi} +
\ald r \left( \frac{\pa}{\pa \xi} \right)_t. \eqno(8)$$
The velocity is given by
$$v =\frac{dr}{dt} = \frac{d}{dt}\left(\frac{\xi}{F} \right)
= - \frac{\xi}{F^2} \ald. \eqno(9)$$
We also substitute
$$B_z (r,t) = B'_z (\xi,t) F^2(t), \eqno(10)$$
$$\Bp (r,t) = \Bp' (\xi,t) F(t). \eqno(11)$$
On substituting (6)--(11) into (4) and (5), a few steps of straightforward
algebra give 
$$\frac{\pa B'_z}{\pa t} 
= F^2 \frac{1}{\xi} \frac{\pa}{\pa \xi} 
\left( \eta \xi \frac{\pa B'_z}{\pa \xi} \right), 
\eqno(12)$$
$$\frac{\pa \Bp'}{\pa t} 
= F^2 \frac{\pa}{\pa \xi} \left[ \eta \frac{1}{\xi} 
\frac{\pa}{\pa \xi}(\xi \Bp') \right].
\eqno(13)$$
It is thus clear that $B'_z(\xi,t)$ and $\Bp' (\xi,t)$ do not change with time
if $\eta = 0$, in accordance with what we expect under the condition of flux
freezing.

The equations (12) and (13) clearly should hold from the centre to the
outer periphery of the flux tube if we assume it to expand self-similarly.

Now we turn to the problem of how to incorporate in our equations the accretion
of poloidal flux to the tube. As the flux tube rises, it collects more poloidal
flux, which gets wrapped around it.  If we are in the frame of the rising flux
tube, it would seem that there is a flow of fluid from the upward direction
bringing the poloidal flux.  In the downward direction, the surrounding fluid
flows away from the flux tube.  However, the tension of the poloidal flux makes
sure that it gets wrapped around the flux tube, after the reconnection in the
wake, as shown in Figure~4 of \citet{Choudhuri:tube.dynamo}.  In our
one-dimensional model, we can approximately take account of this by assuming
that the poloidal flux is brought uniformly from all directions by a radial
inward flow with velocity equal to the velocity with which the fluid is flowing
from the upward direction.  Let us, therefore, consider the nature of flow
velocity from the upward direction. Finding the flow past a cylinder is a
standard problem in incompressible fluid dynamics (which should hold under the
subsonic conditions prevailing in our problem) and is discussed in many
standard textbooks (see, for example, \citen{Choudhuri:book}, \S4.7).  To get
the velocity in the upward direction, we basically have to substitute $\theta =
0$ in the expression (4.54) of \citet{Choudhuri:book} giving the velocity
potential and then differentiate it with respect to $r$. This gives
$$v = - U_{ft} \left( 1 - \frac{\xi_{1/2}^2}{\xi^2} \right), \eqno(14)$$
where $U_{ft}$ is the velocity of rise of the flux tube and $\xi_{1/2}$ is its
radius, defined as the place where $B_z$ falls to half its maximum value (at
the centre of the flux  tube). We assume that there is an isotropic radial
inward flow given by equation (14) towards the flux tube from all directions
{\changes in the region $\xi > \xi_{1/2}$, whereas the flow is zero inside the
flux tube ($\xi <\xi_{1/2}$). }
Such a flow field certainly has a non-zero divergence and other bad
properties.  But it should capture the basic physics of poloidal flux advection
in a one-dimensional model.  We now discuss how we find the advection of the
poloidal flux by such a flow field. The term $\pa (v \Bp)/\pa r$ in equation
(5) gives the advection of the poloidal field by a radial flow.  While deriving
equation (13) from equation (5), we got rid of this term by assuming the
self-similar expansion which gives the velocity field (9).  If there is an
additional velocity field, then the advection term for that velocity would
still persist.  So we put an additional advection term in (13) in our scaled
variable, so that equation (13) becomes
$$\frac{\pa \Bp'}{\pa t} 
= F^2 \frac{\pa}{\pa \xi} \left[ \eta \frac{1}{\xi} 
\frac{\pa}{\pa \xi}(\xi \Bp') \right] - F \frac{\pa}{\pa \xi}(v \Bp').
\eqno(15)$$
The factor $F$ in the last term comes from equation (7).

\subsection{Input parameters: rise of the flux tube}

To understand the magnetic field evolution in the flux tube, we have to solve
equations (12) and (15), starting from some initial configuration and setting 
suitable boundary conditions at the outer periphery of our region of
integration that would allow the free inward advection of the poloidal flux. 
We also need to specify $F$, $\eta$ and $v$. Of these, $v$ has already been
specified by equation (14), although we need to know $U_{ft}$ to get $v$.  We
now discuss how we obtain $F$, $U_{ft}$ and $\eta$ in our model.  We shall come
to a discussion of the initial and boundary conditions later.

As the flux tube rises through the SCZ, we denote its radial distance from the
centre of the Sun by $R$.  The flux tube begins from the bottom of SCZ at
$R=R_b$, where its radius is $\xi_{ft}$ and the external density is
$\rho_{e,0}$. When it rises to $R$ where the external density is $\rho_e$, its
radius becomes $r_{ft}$.  Since the density inside the flux tube would be very
nearly equal to the external density, mass conservation implies
$$R_b \rho_{e,0} \xi_{ft}^2 = R \rho_e r_{ft}^2. $$
From (6) it follows that
$$F = \frac{\xi_{ft}}{r_{ft}} = \sqrt{\frac{R \rho_e}{R_b \rho_{e,0}}}. 
\eqno(16)$$
Thus, to find $F$ as a function of time, we need to find out how $R$ changes as
a function of time and we also require a model of the SCZ which will give us
the value of $\rho_e$ at that value of $R$.  We now discuss how we prescribe a
model for SCZ and how we calculate the rise of the flux tube through this SCZ,
giving $R$ as a function of time.  Since $U_{ft}$ appearing in equation (14) is
essentially given by $dR/dt$, the velocity $v$ as given by equation (14) also
gets completely specified once we know how $R$ varies with $t$. It is also
clear from (16) that $F$ would have a specific value at a certain depth $R$
within the convection zone. The solid line in Figure~1 shows $F$ as a function
of $R$. 

How a horizontal magnetic flux tube rises through SCZ can be studied in a
fairly straightforward fashion (\citen{FMI:horiz.rise}; 
\citen{Choudhuri+Gilman}). We believe that active regions form by the buoyant
rise of a part of a flux tube, while other parts remain anchored at the bottom
of SCZ.  Studies of the rise of such loops show that the upper parts of the
loops move very much like horizontal flux tubes (\citen{FMI:classic};
\citen{Choudhuri:Coriolis};  \citen{DSilva+Choudhuri}). We find out how $R$
varies with $t$ by considering the rise of an axisymmetric flux ring.  The
dynamics of such flux rings has been studied exhaustively by
\citet{Choudhuri+Gilman}.  The forces acting on such a flux ring are (i)
magnetic buoyancy, (ii) the Coriolis force, (iii) magnetic tension, and (iv)
the drag.  The simulations match various aspects of observational data best if
the initial magnetic field at the bottom of SCZ is as strong as $10^5$ G
(\citen{Choudhuri+Gilman}; \citen{Choudhuri:Coriolis};
\citen{DSilva+Choudhuri}; \citen{Fan+:asymm}). For such a strong initial
magnetic field, the Coriolis force is unimportant and we neglect it.  Also,
\citet{Choudhuri+Gilman} made an estimate of magnetic tension compared to
magnetic buoyancy.  Although magnetic tension may be an appreciable fraction of
the magnetic buoyancy at the bottom of SCZ, it becomes negligible in the upper
parts of SCZ.  So we neglect magnetic tension also.  The neglect of the
Coriolis force and magnetic tension makes sure that the flux tube moves
radially and the problem becomes one-dimensional. The only terms in equation
(4) of \citet{Choudhuri+Gilman} which should be of interest to us are
$$2 m_i \frac{d^2 R}{d t^2} = - (m_i - m_e) g_s \left( \frac{\Rs}{R} \right)^2
- \frac{1}{2} C_D \rho_e r_{ft} \left( \frac{dR}{dt} \right)^2, \eqno(17)$$
where $g_s$ is the gravity at the solar surface, whereas $m_i = \pi r_{ft}^2
\rho_i$ and $m_e = \pi r_{ft}^2 \rho_e$ are respectively mass per unit length
of the flux tube and the displaced fluid.  The dimensionless drag coefficient
$C_D$ is found to have a value of around 0.4 in laboratory experiments
(\citen{Goldstein}; \citen{Schlichting}).  Dividing equation (17) by $2 \pi r_{ft}^2
\rho_e$, we get
$$\frac{d^2 R}{d t^2} = \frac{\rho_e - \rho_i}{2 \rho_e} g_s \left( \frac{\Rs}{R} \right)^2
- \frac{ C_D}{4 \pi r_{ft}} \left( \frac{dR}{dt} \right)^2. \eqno(18)$$
It may be noted that $\rho_i \approx \rho_e$ so that we treat them differently
only when we have to consider the difference between them.

\begin{figure}
\includegraphics[width=\columnwidth]{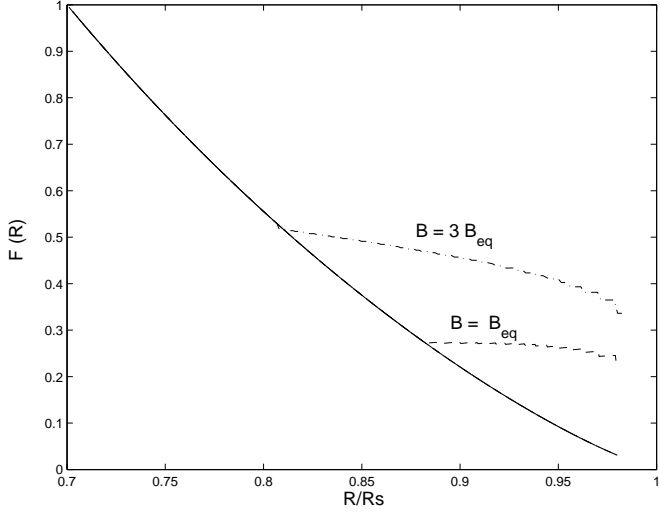}
\caption{ Variation of the expansion factor $F$ with distance $R$ from the centre
of the Sun during the rise of a horizontal flux tube in the convective zone
(modelled as an  adiabatic envelope). Solid: $F$ calculated from eq.~(16). 
Dotted and dashed-dotted: $F$ subject to the requirement $B\geq \Beq$ and
$B\geq 3 \Beq$, respectively.
}
\label{fig:F}
\end{figure}

The magnetic buoyancy factor $(\rho_e - \rho_i)/(2 \rho_e)$ depends on the
background model of SCZ that we use.  If we assume the temperature gradient to
be exactly equal to the adiabatic gradient, then we end up with a polytropic
model for SCZ.  Choudhuri \& Gilman (1987) used a polytropic model with a
choice of parameters which gave a close fit to more detailed models of SCZ. 
The model is described through Equations (10)--(12) of
\citet{Choudhuri+Gilman}, with the values of parameters listed at the beginning
of \S~III.  We use this model of SCZ here.  We also need to use some suitable
thermal condition to calculate the magnetic buoyancy factor $(\rho_e - \rho_i)/
(2 \rho_e)$.  \citet{Choudhuri+Gilman} presented results for three thermal
conditions. We present our calculations here for the simplest case of the flux
tube being in thermal equilibrium with the surrounding, although we have done
some calculations for the other thermal conditions and found the results to be
qualitatively similar.   For the case of thermal equilibrium, the magnetic
buoyancy factor is given by
$$\frac{\rho_e - \rho_i}{2 \rho_e}= \frac{B_0^2}{16 \pi p_{e,0}} 
\left( \frac{T}{T_0} \right)^{(2 - \gamma)/(\gamma - 1)} \left( \frac{R}{R_b} \right)^2
. \eqno(19)$$
Knowing the temperature $T$ at the position $R$ from the SCZ model, we can use
equation (19) to calculate the magnetic buoyancy factor of a flux tube at $R$,
which started with initial magnetic field $B_0$.  Calculating magnetic buoyancy
in this way and using equation (16) to find $r_{ft}$ at the position $R$, we
can integrate equation (18) to find how $R$ changes with time.  From the
variation of $R$ with time, on making use of equation (16), we can find how
$F(t)$ varies with time (Fig.~\ref{fig:F}).  The velocity $U_{ft}$ in equation
(14) is essentially the rise velocity of the flux tube, i.e.
$$U_{ft} = \frac{dR}{dt}. \eqno(20)$$
So, once we know how $R$ changes with $t$, we can use equations (14) and (20)
to obtain $v$ that appears in equation (15).

\subsection{Input parameters: turbulent diffusivity}

As we already pointed out, we need to keep solving equations (12) and (15), as
the flux tube rises, to find out how the magnetic field evolves.  Now that we
know how to find $F(t)$ and $v$ at any time step, we only need to specify the
turbulent diffusivity $\eta$. For this we follow the approach of
\citet{Petrovay+FMI:erosion} and take $\eta$ to be given by the expression
$$\eta = \frac{\eta_0}{1 + |B/\Beq |^{\varkappa}}, \eqno(21)$$  
where $B= \sqrt{B_z^2 + \Bp^2}$ is the amplitude of the magnetic field and
$\Beq $ is the equipartition magnetic field.  We use the convection zone model
of \citet{UKX} to obtain $\Beq $ at different positions $R$ within SCZ.  
Most of our calculations are done by taking $\varkappa = 2$ (except in the
case presented in Figure~3). We
specify $\eta_0$ exactly the same way as \citet{Petrovay+FMI:erosion}.  
If $H$ is the pressure scale height, then we take
$$\eta_0 = \eta_{00} \left( \frac{r_{ft}}{H} \right)^{4/3} \eqno(22)$$
with
$$\eta_{00} = 3 \times 10^{12} \rm{cm}^2 \rm{s}^{-1}. \eqno(23)$$
\citet{Chatterjee2004} found that solar dynamo models give the best fits with
observations for a value of diffusion of this order. It may be noted that we
use equation (22) only when $r_{ft} < H$, which is the case for typical flux tubes
rising through the bulk of SCZ except the uppermost layers.  There we take
$\eta_{0} = \eta_{00}$.  However, even the validity of equation (18) becomes
questionable if $r_{ft} > H$.

Now equations (12) and (15) 
are a set of nonlinear flux-conserving equations that may be solved by an 
explicit two-step Lax-Wendroff scheme. The time step 
$\Delta t$ obeys the stability condition 
$\Delta t < min[(\Delta \xi)^2/\eta_0, \Delta \xi /U_{ft}]$, where $U_{ft}$
is the velocity of rise calculated from equation (18) and $\eta_0$ 
is given by equation (22). 
We use a non-uniform but steady spatial grid with 1500 points which has a finer
resolution of  $\Delta \xi \sim 2.4$ km for $\xi < 3000$ km and increasing
successively by 2$\%$ thereafter upto $\xi = 20,000$ km.

\subsection{Initial and boundary conditions}

Now that we have described how all the various terms appearing in equations
(12) and (15) are specified or can be obtained, we only have to discuss the
initial and boundary conditions used to solve these equations.  Suppose we want
to consider the buoyant rise of a flux tube carrying initial flux $\Psi$ with
an initial magnetic field $B_0$ at the bottom of SCZ. The initial radius of the
flux tube is obviously $\xi_{ft} = \sqrt{\Psi /\pi B_0}$.  We take the initial
condition that $B_z = B_0$ inside $r < \xi_{ft}$ and $B_z = 0$ outside, whereas
$\Bp$ is initially taken to be zero everywhere. The integration region over
which equations (12) and (15) are integrated extends to $r_{\rm out}$ which is
typically taken at $10 \xi_{ft}$.  The solutions are not very sensitive as to
what boundary conditions we use for $B_z'$.  However, since the flux tube keeps
on acquiring poloidal flux which must be advected inward through the boundary
of the integration region, the boundary condition on $\Bp'$ is quite
important.  We assume that the SCZ is filled with a uniform horizontal magnetic
field of 1 G, which gets wrapped around the rising flux tube.  In order to
achieve this, we have to continuously advect $\Bp = 1$ G through the outer
boundary of the integration region.  Looking at equation (11), we realize that
the appropriate boundary condition at $r = r_{\rm out}$ is going to be
$$\Bp' = 1/F(t)\; \; \rm{G} . \eqno(24)$$

\begin{figure}
\includegraphics[width=\columnwidth]{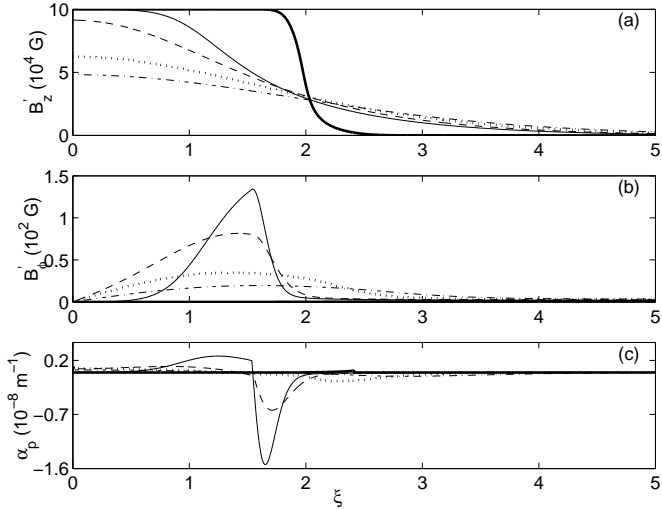}
\caption{Plots of $B_z'$, $\Bp'$ and  $\alpha_p$ as functions of $\xi$ for a
rising flux tube with $\varkappa = 2$ (case A). The different curves correspond to
the profiles of these quantities at the following positions of the flux tube:
$0.7 \Rs$ (thick solid), $0.85\Rs$ (solid), $0.9\Rs$ (dashed), $0.95\Rs$
(dotted), $0.98\Rs$ (dash-dotted).  The flux tube reaches these positions at
times 0 days, 5.3 days,  6.6 days, 7.9 days and 8.7 days after the initial
start. The values of $(B/\Beq )$ at the centres of these flux tubes at these
positions are 10, 1.72, 0.64, 0.098, 0.015.}
\label{fig:fig1}
\end{figure}

\section{Results and discussion}

Flux tube simulations (\citen{Choudhuri+Gilman}; \citen{Choudhuri:Coriolis}; 
\citen{DSilva+Choudhuri}; \citen{Fan+:asymm}) have suggested that theory
matches various aspects of observations best if the flux tubes start with
initial fields of order $10^5$ G at the bottom of SCZ.  We, therefore, carry
out all our calculations by taking an initial magnetic field of $10^5$ G.  The
typical flux carried by a large sunspot is $10^{22}$ Mx.  So we present results
for a flux tube with such flux, which implies that the initial radius at the
bottom of SCZ is $R_b = 1.78 \times 10^3$ km.  We study the rise of the flux
tube by numerically integrating equation (18) with the magnetic buoyancy given
by equation (19).  While the flux tube rises, we study the evolution of the
magnetic field in the flux tube by solving equations (12) and (15) using a
Lax--Wendroff scheme, with $F(t)$ given by equation (16), $v$ by equation (14)
and $\eta$ by equation (21). 

Figures~\ref{fig:fig1} to \ref{fig:fig5} show the magnetic fields of the rising
flux tube at depths $R = 0.7\Rs$, $0.85\Rs$, $0.9\Rs$, $0.95\Rs$, $0.98\Rs$.
Profiles of $B_z'$ as functions of $\xi$ are shown in the top panels, whereas
the middle panels show the profiles of $\Bp'$. As these panels show the
rescaled variables, the actual field strengths can be calculated from them
using Figure~\ref{fig:F} to read off $F$ for each curve, and plugging that value
into equations (10) and (11).  
{\changes Note that $\Bp'$ tends to the value given by (24) for large $\xi$. 
Since this is much smaller than what $\Bp'$ becomes near the flux tube
boundary, it appears in Figures~\ref{fig:fig1} to \ref{fig:fig5} as if $\Bp'$
is going to zero for large $\xi$, although this is not the case.}

The bottom panels provide plots of 
$$\alpha_p= (\nabla \times {\bf B})_z/B_z    \eqno (25) $$
since this is the quantity that essentially all photospheric measurements of
the current helicity actually determine
(\citen{Leka+Skumanich:twist.parameters}; \citen{Burnette+:twist.parameters}). 
It is easy to see that $\alpha_p$, which has the dimension of 1/length, is
invariant to the rescaling of the radial coordinate in our flux tube.

\subsection{Case A: diffuse magnetic field near the surface}

It should be evident from Figure~\ref{fig:fig1} that diffusion does not
play a significant role until the flux tube reaches $0.85 \Rs$.  The diffusion
in the deeper layers turns out to be negligible due to two factors: (i) since
$(B/\Beq )$ is large, the quenching included in equation (21) is quite
efficient; (ii) since $(r_{ft}/H)$ is small, $\eta_0$ as given by equation (22)
turns out to be small. As a result of the low diffusion, the toroidal field
$B_z'$ does not change much, whereas the poloidal flux $\Bp'$ remains confined
in a sheath at the outer periphery of the flux tube (since the low diffusion
does not allow $\Bp'$ to diffuse inward). After the flux tube has risen above
$0.85\Rs$, the toroidal magnetic field inside the flux tube becomes comparable
to the equipartition field $\Beq $.  Then diffusion is able to affect the
magnetic field much more, since $\eta$ given by equation (21) becomes much
larger due to the diminishing role of magnetic suppression of diffusion.  As a
result, $B_z'$ starts diffusing while the flux tube rises through the top
layers of SCZ, whereas $\Bp'$ penetrates to the core of the flux tube instead
of remaining confined to a sheath at the other periphery.

In the very top layers, however, the cross-section of the flux tube becomes
enormous in our model, leading to small values of $F(t)$ and again making
diffusion less important, as can be seen from equations (12) and (15). This is
why we find that the magnetic field profile has not evolved that much from $0.95
\Rs$ to $0.98 \Rs$.  It may be noted that the profile of $B_z'$ no longer has a
sharp edge after the flux tube rises above $0.9 \Rs$, but spreads around.  In
the realistic situation, however, we expect the evolution of $B_z'$ in the
immediate vicinity of the flux tube to be much more complicated.  In the frame
of the flux tube, it would appear that the surrounding fluid is flowing inward
from the upward direction and this would prevent the spread of $B_z'$ in the
upward direction.  On the other hand, the surrounding fluid moves away from the
flux tube in the downward direction and would carry away $B_z'$ with it,
leading to a larger spread of $B_z'$ in the downward direction.  In a
one-dimensional model, the best way of capturing the average behaviour of
$B_z'$ may be not to include any velocity field in the surrounding fluid, as we
have done by not including any advection term in equation (12).  However, an
advection term is included in the evolution equation (15) for $\Bp'$.  Because
of the topology of magnetic field lines, the poloidal field cannot be advected
freely in the downward direction, as can be seen in Figure~4 of 
\citet{Choudhuri:tube.dynamo}.  After a reconnection in the wake, a poloidal
field line should remain wrapped around the flux tube due to its tension. 
Hence the behaviour of the poloidal field is best captured in a one-dimensional
model by including a uniform inward flow from all the directions.  To treat the
evolution of the magnetic field more realistically, it would be necessary to go
beyond one-dimensional models.  Some of us are now involved in developing a
two-dimensional model of this problem.  However, the one-dimensional model
should capture some of the basic effects correct to the right order of
magnitude. 

\subsection{Parameter dependence}

\begin{figure}
\includegraphics[width=\columnwidth]{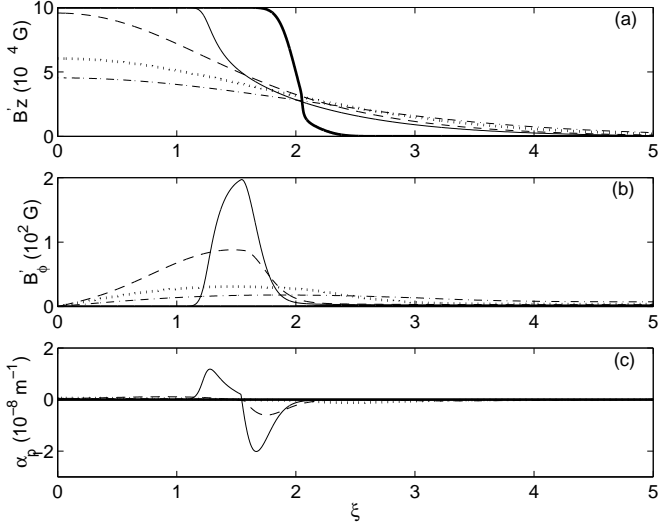}
\caption{Same as Figure~\ref{fig:fig1} but with $\varkappa = 5$. The flux tube
reaches positions 0.0.7$\Rs$, 0.85$\Rs$, 0.9$\Rs$,
0.95$\Rs$,  0.98$\Rs$ at times 0 days, 5.2 days,  6.6 days, 7.9 days and 8.7
days after the initial start.   The values of $(B/\Beq )$ at the centres of
these flux tubes at these positions are 10, 1.74, 0.66, 0.095 and 0.018
respectively.}
\label{fig:fig2}
\end{figure}

\begin{figure}
\includegraphics[width=\columnwidth]{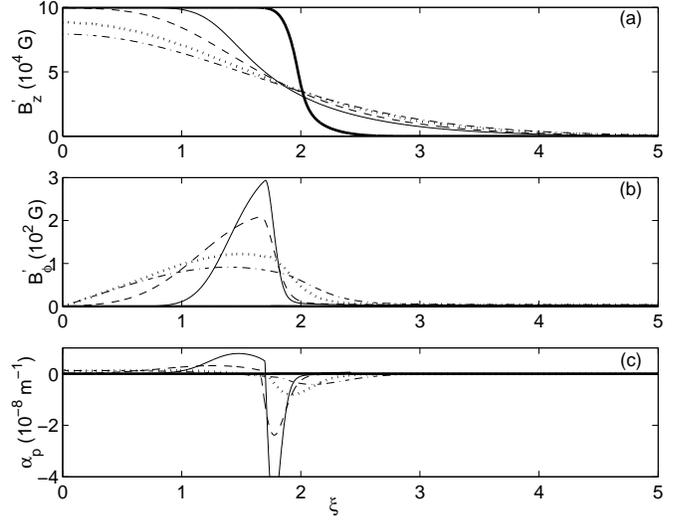}
\caption{Same as Figure~\ref{fig:fig1} except $\eta_{00} = 10^{12}$
cm$^{2}/$s.  The flux tube reaches positions 0.7$\Rs$, 0.85$\Rs$, 0.9$\Rs$,
0.95$\Rs$,  0.98$\Rs$ at times 0 days, 5.2 days,  6.6 days, 7.9 days and 8.7
days after the initial start.   The values of $(B/\Beq )$ at the centres of
these flux tubes at these positions are 10, 1.72, 0.69, 0.14 and 0.03
respectively.}
\label{fig:fig3}
\end{figure}

Since many parameters of the problem are not known well, one very important
question is whether the results presented in Figure~\ref{fig:fig1} are
sufficiently generic or would change substantially on changing the various
parameters. Figure~\ref{fig:fig1} presents results obtained by assuming thermal
equilibrium of the flux tube with the surroundings.  \citet{Choudhuri+Gilman}
presented results for two other thermal conditions: by assuming the interior of
the flux tube as adiabatic with and without a superadiabatic gradient in the
surrounding SCZ. We carried out some calculations with these two thermal
conditions as well.  There are virtually no changes till the flux tube rises
beyond $0.9 \Rs$.  After that, the flux tube rises faster in these two cases
compared to the case of thermal equilibrium and diffusion has less time to
act.  Therefore, we find the effect of diffusion a little bit less when the
flux tube reaches the topmost layers of SCZ.  Figure~\ref{fig:fig2} presents
results obtained by using a quenching index $\varkappa = 5$ instead of $\varkappa =
2$ used in  Figure~\ref{fig:fig2}, whereas all the other things remain the same
as in Figure~\ref{fig:fig1}.  Again the results are not qualitatively
different.  We also study what happens if the diffusivity is made smaller
compared to what is prescribed in (23). Figure~\ref{fig:fig3} presents results
obtained with $\eta_{00}= 10^{12}$ cm$^2$ s$^{-1}$,  whereas all the other
things remain the same as in Figure~\ref{fig:fig1}.  As expected, we find the
diffusion somewhat less. Otherwise, results are not qualitatively different. 
We thus conclude that the results presented in Figure~\ref{fig:fig1} are
sufficiently generic for a reasonable range of parameters, as long as we hold
on to our basic model of the horizontal flux tube rise.

\begin{figure}
\includegraphics[width=\columnwidth]{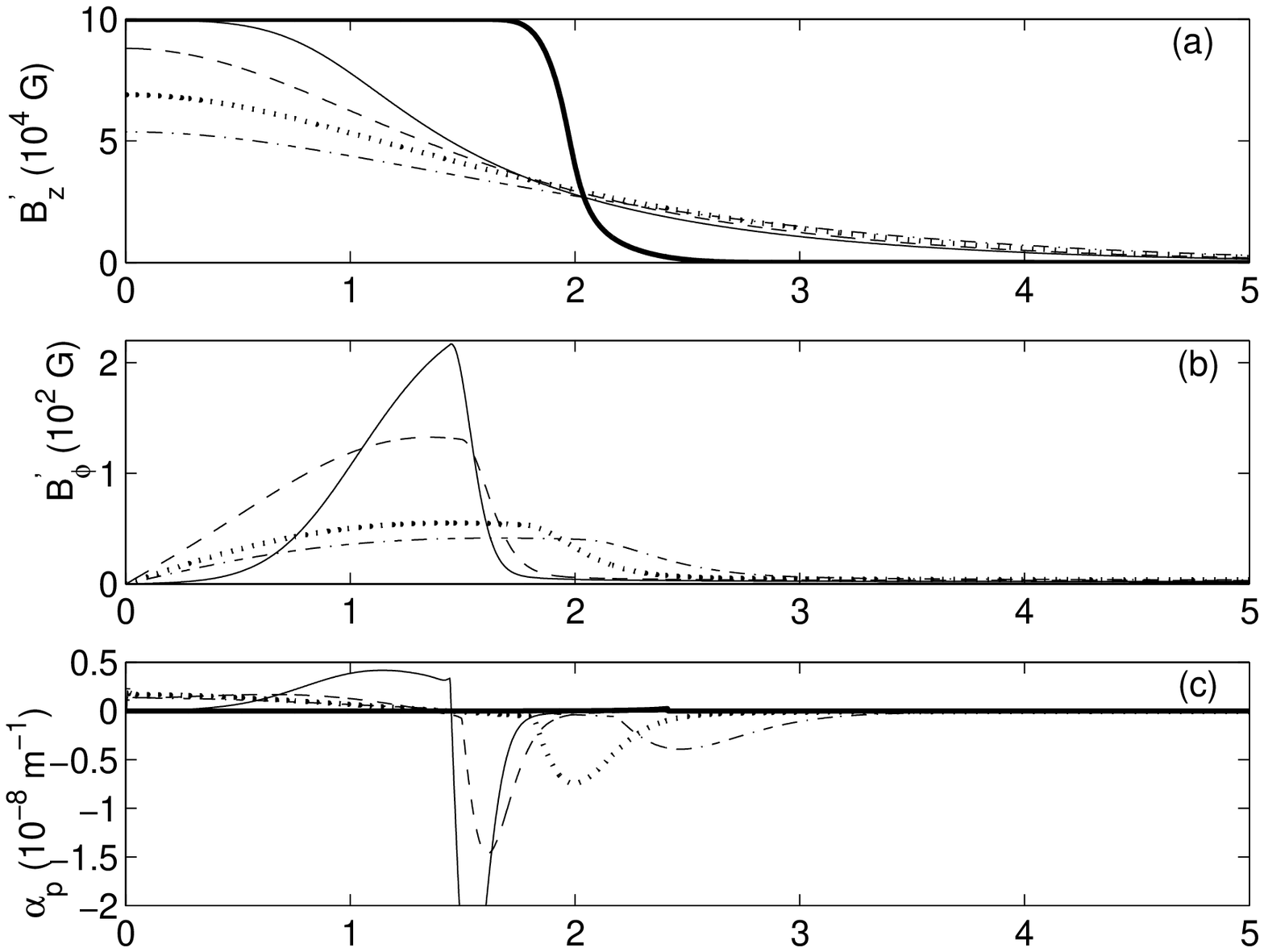}
\caption{Plots of $B_z'$, $\Bp'$ and $\alpha_p$ as functions of $\xi$ for a
rising flux tube.  The field inside the tube is not allowed to decrease below
$\Beq $ (case B1). The different curves correspond to the profiles of these
quantities at the following positions of the flux tube: $0.7 \Rs$ (thick
solid), $0.85 \Rs$ (solid), $0.9\Rs$ (dashed),  $0.95\Rs$ (dotted), $0.98\Rs$
(dash-dotted).  The flux tube reaches these positions at times 0 days, 5.2
days, 6.6 days,  7.7 days and 8.2 days after the initial start.  The values
of $(B/\Beq )$  at the centres of these flux tubes at these positions are 10,
1.72, 1.0, 1.0, 1.0 respectively.}
\label{fig:fig4}
\end{figure}

\begin{figure}
\includegraphics[width=\columnwidth]{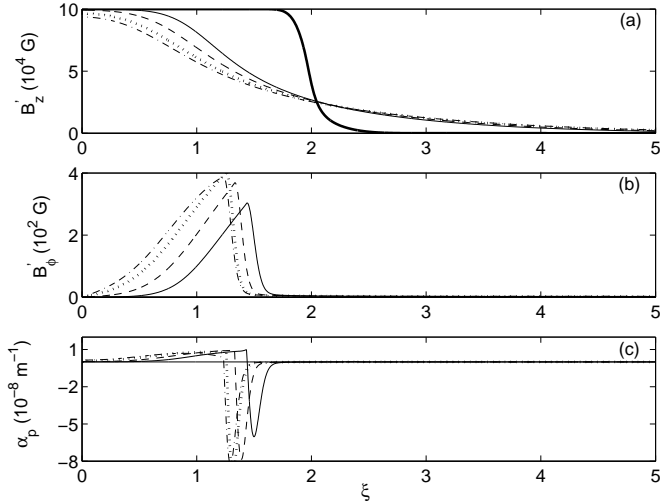} 
\caption{Plots of $B_z'$, $\Bp'$ and $\alpha_p$ as functions of $\xi$ for a
rising flux tube.  The field  inside the flux tube is not allowed to decrease
below $3\Beq $ at any height (case B3). The different curves correspond to the
profiles of these quantities at the following positions of the flux tube: $0.7
\Rs$ (thick solid), $0.85 \Rs$ (solid), $0.9\Rs$ (dashed),  $0.95\Rs$ (dotted),
$0.98\Rs$ (dash-dotted).  The flux tube reaches these positions at times 0
days, 5.2 days,  6.2 days, 6.8 days  and 7.0 days after the initial start.  The
values of $(B/\Beq )$ at the centres of these flux tubes at these positions are
10, 3, 3, 3, 3 respectively.} 
\label{fig:fig5}
\end{figure}

\subsection{Case B: concentrated magnetic field near the surface}

One of the unsatisfactory aspects of the horizontal  flux tube rise model is
that the magnetic field falls to very small values in the top layers of SCZ
when the flux tube expands enormously by moving to a low-density region.  In
the case presented in Figure~\ref{fig:fig1}, our model predicts that the
magnetic field inside the flux tube when it reaches $0.98\,R_\odot$ is about
400 G---an order of magnitude smaller than a typical sunspot field. Since such
a field is much weaker than the equipartition field, it should be clear from
equation (21) that the magnetic quenching of diffusion becomes completely
negligible and diffusion is able to act without being inhibited by the presence
of the magnetic field. The presence of 3000 G magnetic fields in sunspots is a
compelling proof that magnetic fields may never fall to such low values; in
fact, at least in photospheric layers, they remain well above the equipartition
value (the magnetic field inside a typical sunspot being about thrice the
equipartition field).  The non-axisymmetric flux tube simulations
(\citen{Choudhuri:Coriolis}; \citen{DSilva+Choudhuri}; \citen{Fan+:asymm}) show
that the fluid drains from the tops of rising magnetic loops making the
magnetic field there stronger.  Additionally, effects like convective collapse
(\citen{Steiner:Budapest}) can be operative near the solar surface to enhance
the magnetic field.  \citet{Longcope+Choudhuri} have argued that flux tubes get
distorted by convective turbulence in the top layers of SCZ, leading to the
observed scatter of tilt angles around what would be expected from Joy's law. 
Such buffeting of flux tubes by turbulence can also cause an enhancement of
magnetic field by stretching. 

Most of the flux rise simulations are based on the thin flux tube
approximation, which should be valid during the rise of flux tubes to about
$0.9 \Rs$ during which the magnetic field remains sufficiently strong to stay
relatively unaffected by the surrounding turbulence.  So we can presumably
trust the flux rise simulations through the deeper layers of convection.
However, our understanding of what happens during the rise of flux tubes
through the top layers of SCZ is extremely poor.  Techniques of local
helioseismology, such as time-distance seismology, have now evolved to a stage
when the direct study of subsurface structures in emerging active regions has
become possible (\citen{Kosovichev:tomography}; \citen{Hughes+:2AR}). There is
thus hope that this issue may be resolved in the not too distant future. In any
case, the  observation of sunspot magnetic fields indicates that there may be
effects which prevent the magnetic field from falling to values lower than the
equipartition field even in the top layers of SCZ. 

We now present some calculations by artificially not allowing the magnetic
field to fall below the equipartition value.  Suppose the magnetic field in the
interior of the flux tube falls to a value $s \Beq $ at some depth ($s$ being
a numerical factor of the order of unity).  We  assume that the magnetic field
inside the flux tube remains $s \Beq $ in the higher layers as it
rises further.  Here $\Beq $ is the local equipartition value at the
particular depth.  If this is the case, then magnetic buoyancy would be given
by 
$$\frac{\rho_e - \rho_i}{2 \rho_e}= \frac{s^2 \Beq ^2}{16 \pi p_e}  \eqno(26)$$
instead of equation (19).  While we calculate the rise of the flux tube by using
this expression of magnetic buoyancy, we cannot allow the cross-section to
expand indefinitely if the magnetic field has to remain $s \Beq $.  Instead of
equation (16), we calculate $F(t)$ by using the relation
$$F(t) = \sqrt{\frac{s \Beq }{B_0}} \eqno(27)$$
(dashed and dash-dotted lines in Figure~\ref{fig:F}).
By calculating $F(t)$ in this way, we solve equations (12) and (15) to find the
evolution of the magnetic field.  It may be noted that in the expression of
drag in equation (18) also, we have to use $r_{ft} = \xi_{ft}/F(t)$ with $F(t)$
given by equation (27). 

Results for $s=1$ (case B1) and $s=3$ (case B3) are shown in
Figures~\ref{fig:fig4} and \ref{fig:fig5} respectively.  As in
Figures~\ref{fig:fig1}--\ref{fig:fig3}, we plot $B_z'$, $\Bp'$ and $\alpha_p$
as functions of $\xi$ at depths $0.7 \Rs$, $0.85\Rs$, $0.9 \Rs$, $0.95 \Rs$ and
$0.98 \Rs$.  The times taken to reach these depths are given in the figure
captions.  Comparing with the times given in the caption of
Figure~\ref{fig:fig1}, it will be seen that the flux tubes have risen faster
through the top layers of SCZ, since magnetic buoyancy has remained stronger. 
The diffusion remains significantly quenched if the  magnetic field stays
higher than $\Beq $ and has also less time to act because the flux tube rises
faster.  As a result, we see that the effect of diffusion is somewhat less for
case B1 and drastically less for case B3 compared to case A.  We see in
Figure~\ref{fig:fig5} that $B_z'$ has not diffused much and $\Bp'$ has remained
confined in a narrow sheath at the boundary of the flux tube, being unable to
diffuse inward. 

\subsection{Current helicity}

Observations of the current helicity parameter $\alpha_p$ indicate that its
typical average value in an active region is on the order of
$10^{-8}\,$m$^{-1}$ (\citen{Pevtsov1995}; \citen{vDG+:helicity.review};
\citen{Burnette+:twist.parameters}).  Inspecting the lower panels of
Figures~\ref{fig:fig1} to \ref{fig:fig5}  one finds that the typical value of
$\alpha_p$ in the internal parts of the flux tube is of order $\sim 10^{-8}\,$m$^{-1}$
at a depth of $0.85 \Rs$ in all the cases studied.  However, as the flux
reaches the solar surface, in all the cases except the case B3 presented in
Figure~6, the $\Bp$ component spreads out due to diffusion and its gradient
becomes smaller, reducing $\alpha_p$ by about one order of magnitude.
Only if the magnetic field inside the flux tube remains stronger than the
equipartition field (the case B3 represented in 
Figure~6), the $\Bp$ component is unable
to diffuse inside so that its gradient remains strong and $\alpha_p$ is
of order $\sim 10^{-8}\,$m$^{-1}$ even near the surface. 
This suggests that our case B3 may be closest to
reality, i.e. during the rise of the flux tube from $0.9 \Rs$ to $0.98 R_\odot$ 
effective flux concentration processes are at work, keeping the field strength 
at a value somewhat above the equipartition level. This is
consistent with the notion that the main flux concentration effect at work here
is turbulent concentration (i.e. flux expulsion by the turbulent eddies:
\citen{Proctor+Weiss:review}), while convective collapse and thermal relaxation
(\citen{Steiner:Budapest}) are restricted to the shallowest layers above
$0.98\,R_\odot$. 

Note, however, that in the calculations presented here the amplitude and sign
of the poloidal field was assumed not to depend on depth. For alternative
assumptions, significantly different current helicities may result, so the
above conclusion should be treated with proper reservation. Details of the
radial dependence of the poloidal field strength may strongly depend on the
dynamo model.

A more robust feature of the current helicity distributions, present in all
the lower panels of our plots, is the presence of a ring around the tube with a
current helicity of the opposite sense. This is clearly the consequence of the
fact that on the outer side of the accreted sheath the radial gradient of the
azimuthal field, and thus the axial current, is negative. This is an inevitable
corollary of the present mechanism of producing twist in active regions. A
rather strong prediction of this model is, therefore, that a ring of reverse
current helicity should be observed on the periphery of active regions,
somewhere near the edge of the plage.

\subsection{Flux loss from the rising tube}

\begin{figure}
\centering{\includegraphics[width=\columnwidth]{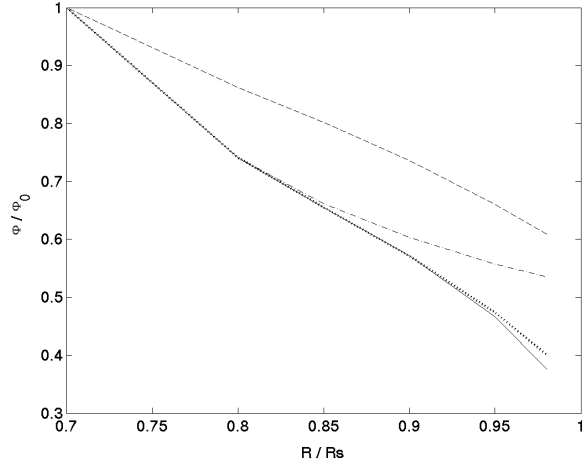}}
\caption{Flux loss as a function of height R inside the convection zone for the
four cases presented in Figure~\ref{fig:fig1} (solid), \ref{fig:fig3}(dashed), 
\ref{fig:fig4}(dotted), \ref{fig:fig5}(dash-dotted).  }
\label{fig:fig6}
\end{figure}

\citet{Petrovay+FMI:erosion} came to the conclusion that a considerable amount
of magnetic flux is lost during the rise of a flux tube through the SCZ (see
their Fig.~10). This conclusion was based on an approximate expression of
inward velocity of turbulent erosion and involved various uncertainties. One
important question is whether our more careful calculations show similar flux
losses. To make an estimate of the flux loss, we keep calculating the flux
$$\psi_{0} = 2 \pi \int_0^{\xi_{0}} B_z'(\xi) \xi d\xi \eqno(28)$$  within the
initial Lagrangian radius $\xi_{0}$ of the flux tube.  Figure~\ref{fig:fig6}
gives plots as $\psi_{0}$ as a function of the position in the SCZ for the
four cases presented in Figures~\ref{fig:fig1}, \ref{fig:fig3}, \ref{fig:fig4}
and \ref{fig:fig5}.  We find that the flux loss is much less if the magnetic
field is not allowed to fall below $\Beq $.  Even in the case when the magnetic
field is allowed to fall to rather low values (the solid line in
Figure~\ref{fig:fig6}), the flux loss is nowhere as substantial as inferred by
\citet{Petrovay+FMI:erosion}. The main reason for this discrepancy is that we
do not make use of the assumption of a continuous ``re-initialization'' of the
decay, as it was done in that paper. The basis of that assumption was that the
flow of external fluid relative to the tube would instantly remove all magnetic
flux lost from the tube. However, in the presence of an azimuthal field
component, field line topology is expected to inhibit such flux removal. We thus
conclude that flux loss during the rise of a flux tube is much less significant
than what \citet{Petrovay+FMI:erosion} found in their  calculations.  This
provides a justification of the flux rise calculations based on the thin flux
tube equation, where it is assumed that the magnetic field is frozen during the
rise of the flux tube. If the above calculations are repeated for the flux tube
rising adiabatically through the convection zone, then the flux losses are
somewhat less. For example, in the case B3, the flux loss is then 41.2\% 
instead of 46.5\% during the isothermal rise.  

\section{Conclusion}

Alternative mechanisms for the origin of twist in active region magnetic fields
include buffeting of the rising flux tubes by helical turbulent motions 
(\citen{Longcope+:Sigma.effect}); the effect of Coriolis force on flows in
rising flux loops (\citen{Fan+Gong:twist}); differential rotation
(\citen{DeVore:diff.rot.helicity}); and helicity generation by the solar dynamo
(\citen{Seehafer+:AdvSpRes}). The relative importance of these processes is
currently a subject of debate
{\changes (\citen{Holder:helicity}). }
It is likely that the accretion of poloidal
fields during the rise of a flux tube is just one contribution to the
development of twist. Its importance may also be reduced by 3D effects:
considering the rise of a finite flux loop instead of an infinite horizontal
tube, the possibility exists for the poloidal field to ``open up'', giving way
to the rising loop with less flux being wrapped around it. It is left for later
multidimensional analyses of this problem to determine the importance of any
such reduction. In any case, the results presented above indicate that the
contribution of poloidal field accretion to the development of twist can be
quite significant, and under favourable circumstances it can potentially
account for most of the current helicity observed in active regions.

In our calculations we found that while the flux tube rises to a depth of about
$0.85 \Rs$, the effect of diffusion is small and the poloidal field remains
confined in a narrow sheath at the periphery of the flux tube.  Afterwards, if
the magnetic field is allowed to fall to very low values as suggested by simple
flux tube rise simulations (case A), then the effect of diffusion is
considerable and the poloidal field is able to penetrate into the interior of
the flux tube. On the other hand, if various effects in the top layers of SCZ
keep the magnetic field above $\Beq $ (case B1) and $3\Beq$ (case B3), then
diffusion is less effective; in case B3 the poloidal field remains confined in
the sheath.  For a poloidal field strength independent of depth, as assumed in
these calculations, the best agreement with the observed current helicity
values is found for case B3. 

One rather strong prediction of our model is the existence of a ring of reverse
current helicity on the periphery of active regions. On the other hand,  the
amplitude of the resulting twist (as measured by the mean current helicity in
the inner parts of the active region) depends sensitively on the assumed
structure (diffuse vs. concentrated/intermittent) of the active region magnetic
field right before its emergence, and on the assumed vertical profile of the
poloidal field. Nevertheless, a mean twist comparable to the observations can
result rather naturally in the model with perhaps the most plausible choice of
assumptions (case B3). 

Thus, by studying the distribution of the azimuthal magnetic field in sunspots
and active regions, and by simultaneously studying the subsurface magnetic
field structure in emerging active regions by means of local helioseismology,
it may be possible to test the present model and to throw some light on the
conditions prevailing during the last phase of flux tube rise through the top
layers of SCZ.   We hope that magnetogram data will improve significantly in
the next few years and it will be possible to draw meaningful inferences. 
Since at present we have very poor understanding of the nature of the magnetic
field or the effect of turbulence when a flux tube rises through layers
immediately below the solar surface, indirect inferences from such observations
of sunspots are of great importance.

{\changes 
Lastly, we have used a simplifying assumption that the poloidal field that
gets wrapped around the flux tube has a constant value of 1 G inside
the SCZ.  Dynamo models like the model 
of Choudhuri et al.\ (2005) can be used to calculate the distribution
of the poloidal field inside the SCZ during different phases of
the solar cycle.  More detailed calculations of helicity can be done 
by using such poloidal field distributions instead of using the simple boundary
condition (24).  Since some groups have started reporting on the possible 
cycle variation of helicity on the basis of the observational data 
(\citen{Bao:helicity}; \citen{Hagino:helicity}), such calculations may 
become relevant in future
when more detailed observational data are available.
}

\acknowledgements

We thank Dibyendu Nandy and an anonymous referee for valuable comments.  This
research was carried out in the framework of the Indo-Hungarian
Inter-Governmental Science \& Technology Cooperation scheme, with the support
of the Hungarian Research \& Technology Innovation Fund and the Department of
Science and Technology of India. K.P. acknowledges support from the UK Particle
Physics and Astronomy Research Council under grant no. PP/X501812/1, from the
ESMN network supported by the European  Commission, and from the OTKA under
grant no.~T043741. P.C. acknowledges financial support from Council for
Scientific and Industrial Research, India under grant no.~9/SPM-20/2005-EMR-I.


\end{document}